\documentclass[preprint,preprintnumbers,superscriptaddress,amsmath]{revtex4}
\usepackage[dvips]{graphics}
\usepackage{epsfig}
\usepackage{psfrag}
\usepackage[psamsfonts]{amssymb}
\usepackage{amsmath}
\usepackage{indentfirst}
\usepackage{amssymb}
\usepackage{url}

\begin{document}

\title{Fertility Heterogeneity as a Mechanism for Power Law Distributions of Recurrence Times}

\author{A. Saichev}
\affiliation{Department of Management, Technology and Economics,
ETH Zurich, Scheuchzerstrasse 7, CH-8092 Zurich, Switzerland}
\affiliation{Mathematical Department,
Nizhny Novgorod State University, Gagarin prosp. 23,
Nizhny Novgorod, 603950, Russia}
\email{saichev@hotmail.com}

\author{D. Sornette}
\affiliation{Department of Management, Technology and Economics,
ETH Zurich, Scheuchzerstrasse 7, CH-8092 Zurich, Switzerland}
\affiliation{ Swiss Finance Institute, 
c/o University of Geneva, 40 blvd. Du Pont d'Arve,
CH 1211 Geneva 4, Switzerland}
\email{dsornette@ethz.ch}

\begin{abstract}

We study the statistical properties of recurrence times in the self-excited
Hawkes conditional Poisson process, the simplest extension of the Poisson
process that takes into account how the past events influence the occurrence
of future events. Specifically, we analyze the impact of the power law 
distribution of fertilities with exponent $\alpha$, where the fertility of an event is the number of 
aftershocks of first generation that it triggers, on the probability
distribution function (pdf) $f(\tau)$ of the recurrence times $\tau$ between successive events.
The other input of the model is an exponential Omori law quantifying
the pdf of waiting times between an event and its first generation aftershocks,
whose characteristic time scale is taken as our time unit.
At short time scales, we discover two intermediate power law asymptotics,
$f(\tau) \sim \tau^{-(2-\alpha)}$ for $\tau \ll \tau_c$ and $f(\tau) \sim \tau^{-\alpha}$
for  $\tau_c \ll \tau \ll 1$, where $\tau_c$ is associated with the self-excited cascades of aftershocks.
For $1 \ll \tau \ll 1/\nu$,  we find a constant plateau $f(\tau) \simeq const$, while
at long times, $1/\nu \lesssim \tau$, $f(\tau) \simeq e^{-\nu \tau}$
has an exponential tail controlled by the arrival rate $\nu$ of exogenous events. 
These results demonstrate a novel mechanism for the generation of 
power laws in the distribution of recurrence times, which results from a
power law distribution of fertilities in the presence of self-excitation and cascades
of triggering.

\end{abstract}

\date{\today}

\maketitle

\pagebreak

\section{Introduction}

The statistics of `recurrence times', defined as the random durations of time intervals 
between two consequent 
events, is widely used to characterize systems punctuated by short-duration occurrences 
interspersed between quiet phases. For instance, 
the statistics of recurrence times between earthquakes is 
the basis for hazard assessment in seismology. The statistics of recurrence times
has recently been the focus of researchers interested in the properties of 
different natural \cite{Zaslavsky91,Corral2003,Corral2004,Davisenetal07} and social systems 
\cite{Barabasi_Nature05,Vasquez_et_al_06}.

The study of recurrence between earthquakes is perhaps the most advanced quantitatively due to the 
availability of data and the high involved stakes.
The statistics of earthquake recurrence times in large geographic domains have been
reported to be characterized by universal intermediate power law asymptotics, 
both for single homogeneous regions 
\cite{Bak2002,Corral2003} and when averaged over multiple regions \cite{Bak2002,Corral2004}. These
intermediate power laws, as well as the scaling properties of
the distribution of recurrence times, were theoretically explained 
by the present authors \cite{SaiSor2006,SaiSor2007}, using the parsimonious 
ETAS model of earthquakes triggering \cite{Ogata88}, which is presently the benchmark model in statistical
seismology. We recall that the acronym ETAS stands for Epidemic-Type Aftershocks
and the ETAS model is an incarnation of the Hawkes self-excited conditional
Poisson process \cite{Hawkes1,Hawkes2,Hawkes3,Hawkes4}. 
Our previous works  \cite{SaiSor2006,SaiSor2007} have 
shown that one does not need to invoke new universal laws or fancy scaling
in order to explain quantitatively with high accuracy the previously 
reported scaling laws \cite{Bak2002,Corral2003,Bak2002,Corral2004}. 
In other words, the finding reported in \cite{Bak2002,Corral2003,Corral2004} do not contain 
evidence for any new physics/geophysics laws but constitute just
a reformulation of the following laws:
\begin{itemize}
\item  Earthquakes tend to trigger other earthquakes according
to the same triggering mechanism,  independently of their magnitudes.

\item The Omori-Utsu law for aftershocks, generalized into the phenomenon of earthquake
triggering where all earthquakes are treated on the same footing, states that
the rate of events that are triggered by a preceding event that occurred at time $0$ decays as 
\begin{equation}\label{omorilawexpr}
f_1(t) = \frac{\theta t_0^\theta}{(t_0+t)^{1+\theta}} , \qquad 0<\theta \ll 1 , \qquad t_0>0 , \qquad t> 0~.
\end{equation}
The function $f_1(t)$ can also be interpreted as the probability density function (pdf) of
the durations of the waiting time intervals between 
the reference ``mother'' event and the triggered events of first generation
corresponding to direct triggered by the mother event. The constant $t_0$ 
describes a characteristic microscopic time scale of the generalized Omori law that
ensures regularization at small time and normalization.
\end{itemize}

Our previous analytical derivations \cite{SaiSor2006,SaiSor2007}, found in excellent agreement with
empirical data \cite{Bak2002,Corral2003,Corral2004}, was essentially based on the 
long-memory of the Omori law (\ref{omorilawexpr}), 
$f_1(t) \sim t^{-1-\theta}$ for large $t$ with $0 \leq \theta <1$.
However, it did not take into account the impact of heterogenous fertilities,
which come in wildly varying values.
Indeed, the number of daughters triggered by an earthquake
of a given magnitude grows exponentially with its magnitude.
For instance, a magnitude 8-earthquake
may have tens of thousands of aftershocks of magnitude larger than 2 while
a magnitude 2-earthquake may generate no more than 0.1 earthquake on average
of magnitude larger than 2 \cite{Helmstteterfertility}. Given the fact that the
distribution of magnitudes is itself an exponentially decaying function of magnitudes
(called the Gutenberg-Richter law), this translates into a
heavy tail distribution of fertilities \cite{Saihelmsor05}, i.e.
the distribution of the number of first generation events 
triggered by a given event has the following power law asymptotic:
\begin{equation}\label{poneralasymp}
p_1(r) \sim r^{-\alpha-1} , \qquad r\to \infty , \qquad \alpha \in (1,2) ~.
\end{equation}
Precisely, $p_1(r)$ is the probability that the random number $R_1$ 
of first generation aftershocks triggered independently 
by a given mother event is equal to a given integer $r$.

In fact, the main approximation in our previous work \cite{SaiSor2006,SaiSor2007}
was to consider that, for the estimation 
of the distribution of recurrence times, it is sufficient to assume that each mother event
triggers at most one event, so that the power law \eqref{poneralasymp} is completely irrelevant.
This surprising approximation was justified by the focus on the tail of the distribution
of recurrence times, for which typically only one event, among the
set of events triggered by a given earthquake, does contribute.

The goal of the present paper is to reexamine this approximation and 
present an exact analysis of the impact of 
the power law form \eqref{poneralasymp} of the distribution of 
fertilities on the distribution of the recurrence times.
To make the analysis feasible and exact, we consider the case where the Omori law
is no more heavy-tailed but has a shorter memory in the form of an exponential distribution, 
expressed by a suitable choice of time units in the form 
\begin{equation}\label{expdisdef}
f_1(t) = e^{-t} .
\end{equation}
In addition to getting exact analytical expressions, our study of the case
of an exponential memory kernel (\ref{expdisdef}) is motivated by the fact that, for
many applications, this is the default assumption
 \cite{Chavezetal05,BauwensHautsch09,Eymanetal10,Azizetal10,Aitsahaliaetal10,SalmonTham08,Filisor12}.
Therefore, this parameterization has an genuine interest and intrinsic value.
This exponential memory function should not be confused with the Poisson model, which has
no memory. In contrast, the Hawkes process takes into account the full
set of interactions between all past events and the future events, mediated
by the influence function given by $f_1(t)$.

The main result of the present paper is the exact expression for the full distribution $f(t)$ of recurrence times
that results from all the possible cascades of triggering of events over all generations.
We make explicit the substantial dependence of the power law exponents
characterizing the distribution $f(t)$ on the exponent $\alpha$ 
of the power law tail \eqref{poneralasymp} for the distribution of fertilities 
and on the branching ratio $n$,  defined as the mean number of events of first
generation triggered per event: 
\begin{equation}
n= \text{E}\left[R_1\right] ~.
\label{wryjujiuk}
\end{equation}
 
The paper is organized as follows. Section 2 present the self-excited conditional
Hawkes Poisson process and the main exact equations obtained using 
generating probability functions. Section 3 studies the probability of quiescence
within a given fixed time interval and derives different statistical properties of earthquake clusters, such as the
mean number of earthquakes in clusters and their mean duration.
Section 4 presents the main results concerning the probability density functions (pdf)
of the waiting times between successive earthquakes. Section 5 summarizes 
our main results and concludes. 
The Appendix makes more specific the analytical model used in our derivations
and describes useful statistical properties of first generation aftershocks.
While we use the language of seismology and events are named `earthquakes', 
our results obviously apply to the many natural 
and social-economic-financial systems in which self-excitation occurs.

 \section{Branching model of earthquake triggering}

Before discussing the statistical properties of recurrence times, 
let us develop the statistical description of the random number of earthquakes
occurring within the time window $(t,t+\tau)$.
The Hawkes model that we consider assumes that there are 
exogenous events (called ``immigrants'' in the literature on branching processes, 
or ``noise events''), occurring spontaneously according to 
a Poissonian stationary flow statistics. Thus, successive instants
$\dots<t_{-1}<t_0 < t_1 < t_2 < \dots$
of the noise earthquakes belong to the stationary Poisson point process
with mean rate $\nu=\text{const}$. Then, the Hawkes model 
assumes that any given noise earthquake, occurring at $\{t_k\}$, triggers 
a total of $R_1^k$ first generation earthquakes. We assume that 
the set of the total numbers $\{R_1^k\}$ of first generation aftershocks triggered
by noise earthquakes are iid random integers, with the same generating probability 
function (GPF) $G_1(z)$. Furthermore, the distribution of waiting times $t$ between the time $t_k$
of a given noise earthquake and the occurrences of its aftershocks is assumed to be $f_1(t)$
defined by expression (\ref{expdisdef}).
In turn, each of the aftershocks of first generation triggered by a given 
noise earthquake also triggers independently its own first generation aftershocks, 
with the same statistical properties (same GPF) as the first generation aftershocks.
 
Specifically, the Hawkes self-excited conditional Poisson process
is defined by the following form of the Intensity function
\begin{equation}
\lambda(t | H_t, {\cal P}) = \lambda_{\rm noise}(t) + \sum_{i | t_{i} < t} R_1^i~  f(t-t_i)~,
\label{hyjuetg2tgj}
\end{equation}
where the history $H_t = \{ t_i \}_{1 \leq i \leq i_t,~ t_{i_t} \leq t < t_{i_t+1} }$ includes all events
that occurred before the present time $t$ and the sum
in expression (\ref{hyjuetg2tgj}) runs over all past triggered events.  The set of parameters is 
denoted by the symbol ${\cal P}$. The term $\lambda_{\rm noise}(t)$ means
that there are some external noise (or immigrant, exogenous)
sources occurring according to a Poisson process
with intensity $\lambda_{\rm noise}(t)$, which may be a function of time, but all other events can be both triggered by previous events and can themselves trigger their offsprings.  This gives rise
to the existence of many generations of events. In the sequel, 
we will consider only the case where $\lambda_{\rm noise}(t) = \lambda_{\rm noise}$ is constant.

Under the above definitions and assumptions,
one can prove that the GPF of the total number of all earthquakes 
(including noise and triggered events of all generations)
occurring within the time window $(t,t+\tau)$  can be expressed as a product
of two terms:
\begin{equation}\label{thztauexpr}
\Theta(z;\tau) =\Theta_\text{out}(z;\tau) \cdot \Theta_\text{in}(z;\tau) ~.
\end{equation}
The first term $\Theta_\text{out}(z;\tau)$ is the GPF of the number of all earthquakes
in $(t,t+\tau)$ triggered by noise and triggered earthquakes that occurred up to time $t$.
The second term $\Theta_\text{in}(z;\tau)$ is the GPF of the number of all noise earthquakes
that occurred within $(t,t+\tau)$ and of all earthquakes triggered in that window
by events also in $(t,t+\tau)$. The factorization of $\Theta(z;\tau)$ given by
expression (\ref{thztauexpr}) simply expresses the independence between
the branching processes starting outside and within the time window $(t,t+\tau)$.

We have previously shown \cite{SaiSor2007,SaiSor2006a}) that 
$\Theta_\text{out}(z;\tau)$ and $\Theta_\text{in}(z;\tau)$ are given respectively by
\begin{equation}\label{mathadef}
\Theta_\text{out}(z;\tau) = \exp\left(\nu \int_0^\infty \left[ \mathcal{G}(z;t,\tau) -1 \right] dt \right)~,
\end{equation}
and
\begin{equation}\label{mathbdef}
\Theta_\text{in}(z;\tau) = \exp\left(\nu \int_0^\tau \left[ z G(z;\tau)-1\right] dt \right)~ ,
\end{equation}
where the functions $\mathcal{G}(z;t,\tau)$ and $G(z;\tau)$ satisfy the following nonlinear integral equations:
\begin{equation}\label{gztallaft}
\begin{array}{c} \displaystyle
G(z;\tau) = Q\left[H(z;\tau)\right] ~,
\\[1mm]\displaystyle
H(z;\tau) = \rho(t) - z f_1(\tau) \otimes G(z;\tau)~ ,
\\[1mm]\displaystyle
G(z;0) = 1 , \qquad H(z;0) = 0 ~.
\end{array}
\end{equation}
and
\begin{equation}\label{gztallaftseq}
\begin{array}{c} \displaystyle
\mathcal{G}(z;t,\tau) = Q[\mathcal{H}(z;t,\tau)] ~,
\\[1mm]\displaystyle
\mathcal{H}(z;t,\tau) = \rho(t+\tau)  - \mathcal{G}(z;t,\tau) \otimes f_1(t) - z G(z;\tau) \otimes f_1(t+\tau)~ ,
\\[1mm]\displaystyle
\mathcal{G}(z;0,\tau) = G(z;\tau) , \qquad \mathcal{H}(z;0,\tau) = H(z;\tau)~ .
\end{array}
\end{equation}
The symbol $\otimes$ represents the convolution operator with respect to the repeating time arguments $t$ or $\tau$. We have introduced the auxiliary function
\begin{equation}\label{qytrugoneomz}
Q(y) := G_1(1-y)~ ,
\end{equation}
and the cumulative distribution function (cdf) of the first generation aftershocks instants
\begin{equation}
\rho(t) = \int_0^t f_1(t') dt'~ .
\label{syjiukoiloyi}
\end{equation}

\begin{figure}
\begin{center}
  \includegraphics[width=0.8\linewidth]{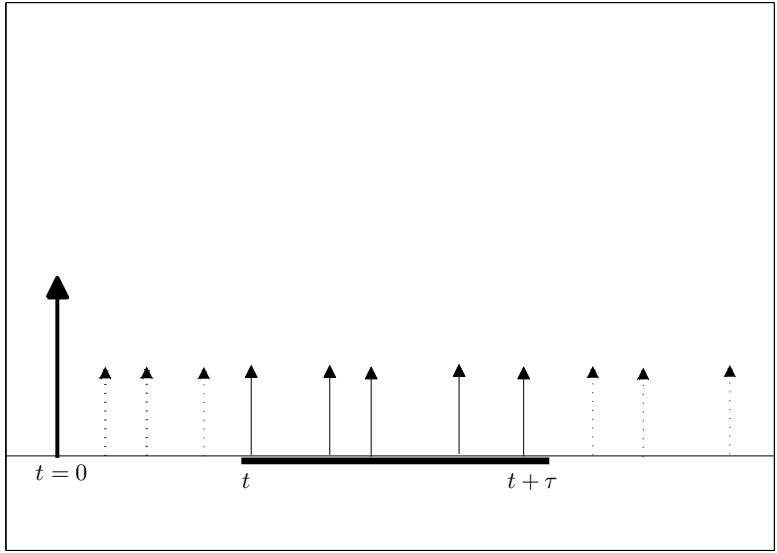}\\
\end{center}
  \caption{Schematic illustration of the geometric sense of function $\mathcal{G}(z;t,\tau)$. It is the GPF of the random number $R(t,\tau)$ of all generations aftershocks within the window $(t,t+\tau)$
triggered by a noise earthquake that occurred at the origin of time $t=0$ (large solid arrow). 
The aftershocks within $(t,t+\tau)$ are depicted by the small solid arrows. The aftershocks triggered outside the window $(t,t+\tau)$ are depicted by the small dotted arrows. In the picture, $R(t,\tau)=5$.}
\label{gztgeom}
\end{figure}

The functions $G(z;\tau)$ and $\mathcal{G}(z;t,\tau)$ have 
an intuitive geometric sense, illustrated by figure~\ref{gztgeom}. $G(z;\tau)$ is the GPF of the random number of 
aftershocks of all generations triggered till the current time $t=\tau$ by some noise earthquake (or some aftershock) that occurred at the origin of time $t=0$. 
In turn, $\mathcal{G}(z;t,\tau)$ is the GPF of the 
random number of all generations aftershocks triggered within the window $(t,t+\tau)$ (for $t>0$) 
by some noise earthquake that occurred at the origin of time. 
The GPFs $G(z;\tau)$ and $\mathcal{G}(z;t,\tau)$ are related by their definition as follows:
\begin{equation}
G(z;0,\tau) = G(z;\tau) .
\end{equation}

A key input of the theory within the formalism of GPF is thus $G_1(y)$ or $Q(y)$ (via its
definition (\ref{qytrugoneomz})). Given expression (\ref{poneralasymp}),
we show in the Appendix that a convenient form is
\begin{equation}\label{gonepowgamdef}
G_1(z) = 1 - n (1-z) + \kappa (1-z)^\alpha , \qquad \alpha\in(1,2] ~,
\end{equation}
so that the corresponding auxiliary function $Q(y)$ \eqref{qytrugoneomz} takes the form
\begin{equation}\label{qyalphadef}
Q(y) = 1 - n y + \kappa y^\alpha~ .
\end{equation}

\section{Probability of quiescence, mean number of earthquakes and mean duration of clusters}

\subsection{Probability of quiescence}

For $z=0$, relation \eqref{thztauexpr} reduces to
\begin{equation}\label{probsztauexpr}
\text{P}(\tau) =\text{P}_\text{out}(\tau) \cdot \text{P}_\text{in}(\tau) ,
\end{equation}
where
\begin{equation}
\text{P}(\tau) = \Theta(z=0;\tau)
\end{equation}
is the probability that there are no earthquakes (including the noise earthquakes and their
aftershocks of all generations) within the window $(t,t+\tau)$. 
$\text{P}(\tau)$ can be decomposed as the product of 
the probability  $\text{P}_\text{in}(\tau)$ that no noise earthquakes
occur within $(t,t+\tau)$ and the probability $\text{P}_\text{out}(\tau)$
that no aftershocks occur within  $(t,t+\tau)$ that could have been triggered
by noise earthquakes and their aftershocks that occurred before and until time $t$.

In the general case, $\text{P}_\text{in}(\tau)$ is given by
\begin{equation}\label{prtauin}
\text{P}_\text{in}(\tau) = e^{-\nu \tau} ~.
\end{equation}
In turn, $\text{P}_\text{out}(\tau) = \Theta_\text{out}(z=0;\tau)$ by definition. 
Using relation \eqref{mathadef} and equations \eqref{gztallaft}, \eqref{gztallaftseq},
we obtain
\begin{equation}\label{protauout}
\text{P}_\text{out}(\tau) = \exp\left(\nu \int_0^\infty \left[ \mathcal{G}(t,\tau) -1 \right] dt \right) ~,
\end{equation}
where
\begin{equation}
\mathcal{G}(t,\tau) = \mathcal{G}(z=0;t,\tau)
\end{equation}
is defined by 
\begin{equation}\label{gtauoutexpr}
\mathcal{G}(t,\tau) = Q[\mathcal{H}(t,\tau)]~ .
\end{equation}
The auxiliary function $\mathcal{H}(t,\tau)$ is solution of the nonlinear integral equation
\begin{equation}\label{gtouteqs}
\begin{array}{c} \displaystyle
\mathcal{H}(t,\tau) = \rho(t+\tau)  - Q[\mathcal{H}(t,\tau)] \otimes f_1(t) ~,
\\[1mm]\displaystyle
\mathcal{H}(0,\tau) = \rho(\tau)~ .
\end{array}
\end{equation}

\subsection{Solution of equation \eqref{gtouteqs}  and determination of $\text{P}(\tau)$
for the exponential pdf $f_1(t)$}

Using the exponential form of the Omori law (\ref{expdisdef}),  it is possible to 
obtain an exact analytical solution of equation \eqref{gtouteqs}, 
which gives us the possibility to explore in detail the probabilistic properties of recurrence times.

Using the form (\ref{expdisdef}), it is easy to show that equation \eqref{gtouteqs} reduces to the initial value problem
\begin{equation}
\frac{d \mathcal{H}}{dt} + \mathcal{H} + Q\left[\mathcal{H}\right] = 1 , \qquad \mathcal{H}(0,\tau) = \rho(\tau) ~.
\end{equation}
Using expression \eqref{qyalphadef} for $Q(y)$ leads to
\begin{equation}\label{matheqalp}
\frac{d \mathcal{H}}{dt} + (1-n) \mathcal{H} +\kappa \mathcal{H}^\alpha = 0 , \qquad \mathcal{H}(0,\tau) = \rho(\tau) ~,
\end{equation}
whose solution is given by
\begin{equation}\label{mathxtexpr}
\mathcal{H}(t,\tau) =
\left[ (1 - e^{-\tau})^{1-\alpha} ~e^{(\alpha-1)(1-n)t} + \gamma \left(e^{(\alpha-1)(1-n)t} - 1\right) \right]^{1/(1-\alpha)}~,
\end{equation}
where
\begin{equation}\label{gamtilaldef}
\gamma = \frac{\kappa}{1-n} ~.
\end{equation}
We have used the fact that 
\begin{equation}\label{atauexp}
\rho(\tau) = 1 - e^{-\tau}~,
\end{equation}
as derived from the exponential pdf $f_1(t)$ given by \eqref{expdisdef}
and definition (\ref{syjiukoiloyi}).

We can now rewrite probability $\text{P}_\text{out}(\tau)$ \eqref{protauout} in the form
\begin{equation}\label{mathafoverline}
\text{P}_\text{out}(\tau) = e^{-\nu \overline{F}(\tau)}~ ,
\end{equation}
where
\begin{equation}\label{overfdef}
\overline{F}(\tau) = \int_0^\infty \left[1- \mathcal{G}(t,\tau) \right] dt~.
\end{equation}
Taking into account expression \eqref{mathxtexpr} and the equality
\begin{equation}
\mathcal{G}(t,\tau) = Q[\mathcal{H}(t,\tau)] = 1 - n \mathcal{H} +\kappa \mathcal{H}^\alpha ,
\end{equation}
the explicit calculation of integral \eqref{overfdef} yields
\begin{equation}\label{overfexpr}
\begin{array}{c} \displaystyle
\overline{F}(\tau) = \overline{F}(n,\kappa,\alpha,\rho) =
\\[2mm] \displaystyle
\frac{\gamma^{1/(1-\alpha)}}{(\alpha-1) (1-n)}
\bigg[ n B\left(\frac{\gamma \rho^{\alpha}}{\rho+\gamma \rho^{\alpha}}, \frac{1}{\alpha-1}, \frac{\alpha-2}{\alpha-1}\right) -
\\[4mm] \displaystyle
(1-n) B\left(\frac{\gamma \rho^{\alpha}}{\rho+\gamma \rho^{\alpha}}, \frac{\alpha}{\alpha-1}, \frac{1}{1-\alpha}\right)  \bigg] ~,
\end{array}
\end{equation}
where $\rho=\rho(\tau)$ and $B(x;a,b)$ is the incomplete beta function
\begin{equation}
B(x;a,b) = \int_0^x s^{a-1} (1-s)^{b-1} ds~ .
\end{equation}

In view of the key role played by the function $\overline{F}(n,\kappa,\alpha,\rho)$ in the following, 
it is useful to describe some of its properties. 
A first result of interest is its limit behavior as the branching ratio $n$ tends to $1$.
Recall that this limit corresponds to the critical regime of the Hawkes process,
separating the subcritical phase $n<1$ and the supercritical phase $n>1$.
For $n<1$, each noise earthquake has only a finite number of aftershocks.
For $n>1$, there is a non-zero probability that a single noise earthquake
may generate an infinite number and infinitely long-lives sequence of aftershocks.
The relevant physical regime is thus $n \leq 1$ and the boundary value $n=1$
plays a special role, especially when one remembers that $n$ can also be interpreted
as the ratio of the total number of triggered events to the total number of events \cite{HelmstteterSornette03}.
Hence, when $n \to 1$, most of the observed activity is endogenous, i.e., triggered
by past activity.

Thus, the limit of $\overline{F}(n,\kappa,\alpha,\rho)$ as $n \to 1$ reads
\begin{equation}\label{overefeneqone}
\overline{F}(\kappa,\alpha,\rho) = \lim_{n\to 1} \overline{F}(n,\kappa,\alpha,\rho) = \frac{\rho^{2-\alpha}}{\kappa (2-\alpha)} - \rho , \quad 1<\alpha < 2 .
\end{equation}

For $n <1$, it is convenient to choose values of $\alpha$ that take the form
\begin{equation}
\alpha = 1 +\frac{1}{m} ~,
\end{equation}
so that it is possible to express the function
\begin{equation}
\mathcal{F}_m(n,\kappa,\rho) = \overline{F}\left(n,\kappa,1+\frac{1}{m},\rho\right)
\end{equation}
under the form
\begin{equation}\label{mathefexpr}
\begin{array}{c} \displaystyle
\mathcal{F}_m(n,\kappa,\rho) = \frac{n \rho}{1-n} +
\\[4mm] \displaystyle
\frac{m}{\kappa^m} (n-1)^{m-1} \left[ \ln\left(1+\frac{\kappa}{1-n} \rho^{1/m} \right) - \ln_m\left(1+\frac{\kappa}{1-n} \rho^{1/m} \right) \right]~ .
\end{array}
\end{equation}
The auxiliary function $\ln_m(1+x)$ is defined as the sum of the first $m$ terms of 
the Taylor series expansion of  the logarithm function $\ln(1+x)$ with respect to $x$:
\begin{equation}
\ln_m(1+x) = - \sum_{k=1}^m \frac{(-x)^k}{k}~ .
\end{equation}

For $m=1$ ($\alpha=2$), we have
\begin{equation}\label{overfaltone}
\mathcal{F}_1(n,\kappa,\rho) = \frac{1}{\kappa} \ln\left[1 + \frac{\kappa \rho}{1-n} \right ] - \rho ~.
\end{equation}
For $m=2$ ($\alpha=3/2$), we have
\begin{equation}\label{overfaltwo}
\begin{array}{c} \displaystyle
\mathcal{F}_2(n,\kappa,\rho) = \frac{2}{\kappa} \sqrt{\rho} - \rho - \frac{2}{\kappa^2} (1-n) \ln\left(1+ \frac{\kappa \sqrt{\rho} }{1-n} \right)~.
\end{array}
\end{equation}
For $m=3$ ($\alpha=4/3$), we have
\begin{equation}\label{overf43}
\begin{array}{c} \displaystyle
\mathcal{F}_3(n,\kappa,\rho)= -\frac{3}{\kappa^2}(1-n) \rho^{1/3} + \frac{3}{2\kappa}~  \rho^{2/3}- \rho +
\\[4mm] \displaystyle
\frac{3}{\kappa^3} (1-n)^2 \ln\left(1+ \frac{\kappa \rho^{1/3}}{1-n} ~ \right)~.
\end{array}
\end{equation}

\subsection{Mean duration of seismic clusters}

\subsubsection{Seismic clusters of all types}

Let us consider the random duration $\Delta_k$ of the aftershock cluster triggered by 
the $k$th noise earthquake that occurred at time $t_k$. By definition, 
\begin{equation}
\Delta_k =t_\text{last}^k- t_k ~,
\end{equation}
where $t_\text{last}^k$ is the occurrence time of the last of its triggered aftershocks
over all generations. The mean value of $\Delta_k$ is by definition
$\langle \Delta \rangle = \int_0^\infty \varrho \cdot w(\varrho) d\varrho$, 
where $w(\varrho)$ is the pdf of the random cluster durations $\{\Delta_k\}$. 

By hypothesis, the instants $\{t_k\}$ of the noise earthquakes form a Poissonian point process
with mean rate $\nu$. Moreover, within the 
Hawkes branching process model, the clusters durations $\{\Delta_k\}$ 
are iid random variables. One can easily show 
that the probability $\text{P}_\text{out}(\tau)$ given by  \eqref{mathafoverline},
that no aftershocks occur within  $(t,t+\tau)$ that could have been triggered
by noise earthquakes and their aftershocks that occurred before and until time $t$,
take the following value in the limit $\tau \to +\infty$:
\begin{equation}\label{pinfdel}
\text{P}_\text{out}(\infty)  = e^{-\nu \langle \Delta \rangle}~,
\end{equation}
Thus, $\text{P}_\text{out}(\infty)$ is the probability that
all noise earthquakes that occurred up to time $t$ do not trigger any 
aftershock after $t$.
It follows from relation \eqref{mathafoverline} and (\ref{pinfdel}) that
\begin{equation}\label{nglDelgen}
\langle \Delta \rangle = \overline{F}_\infty(n,\kappa,\alpha, \rho=1)~ .
\end{equation}

\begin{figure}
\begin{center}
  \includegraphics[width=0.8\linewidth]{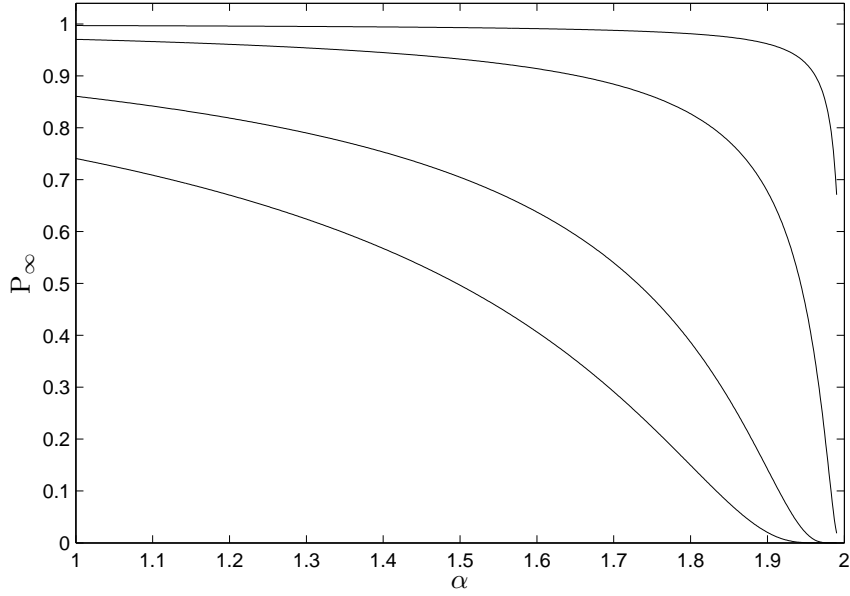}\\
\end{center}
  \caption{Dependence with respect to the exponent $\alpha$ 
of the probability $\text{P}_\text{out}(\infty)$ \eqref{pinfdel} 
that no aftershocks occur within  $(t,t+\tau)$ that could have been triggered
by noise earthquakes and their aftershocks that occurred before and until time $t$.
The plot corresponds to the critical case $n=1$ with $\kappa=0.25$. 
Bottom to top: $\nu=0.1;0.05; 0.01; 0.001$.}\label{pinfcrit}
\end{figure}
 
Now, $\overline{F}_\infty(n,\kappa,\alpha, \rho=1)$ can be obtained
as the value of the function $\overline{F}(\tau)$ given by \eqref{overfexpr} at $\tau \to +\infty$:
\begin{equation}\label{finfdef}
\overline{F}_\infty(n,\kappa,\alpha, \rho) = \overline{F}(\infty) ~.
\end{equation}
We obtain
\begin{equation}
\begin{array}{c} \displaystyle
\overline{F}(n,\kappa,\alpha,\rho=1) =
\\[4mm] \displaystyle
\frac{\gamma^{1/(1-\alpha)}}{(\alpha-1) (1-n)}
\bigg[ n B\left(\frac{\kappa}{\kappa+1-n}, \frac{1}{\alpha-1}, \frac{\alpha-2}{\alpha-1}\right) -
\\[4mm] \displaystyle
(1-n) B\left(\frac{\kappa}{\kappa+1-n}, \frac{\alpha}{\alpha-1}, \frac{1}{1-\alpha}\right)  \bigg]~ .
\end{array}
\label{hyjruiu5jyne}
\end{equation}
In particular, in the critical case $n=1$, we have
\begin{equation}
\overline{F}(1,\kappa,\alpha,1) = \frac{1}{\kappa(2-\alpha)} - 1~ ,
\end{equation}
and thus
\begin{equation}\label{angledenone}
\langle \Delta \rangle = \frac{1}{\kappa (2-\alpha)} - 1 , \qquad n= 1 , \qquad \kappa < \alpha^{-1}~ .
\end{equation}

Figure~\ref{pinfcrit} shows the probability $\text{P}_\text{out}(\infty)$  \eqref{pinfdel} as
a function of the exponent $\alpha$, in the critical case $n=1$.

\subsubsection{Seismic clusters with at least $m \geq 1$ aftershock}

The mean duration $\langle \Delta \rangle$ of clusters 
given by expression \eqref{nglDelgen} with (\ref{hyjruiu5jyne})
includes the contribution of the empty clusters for which the noise earthquake 
does not trigger any aftershock.
It is thus interesting to evaluate another derived quantity $\langle\Delta^1\rangle$, 
defined as the mean duration of clusters
that contain at least one aftershock. To get  $\langle\Delta^1\rangle$,
we  divide  $\langle \Delta \rangle$ by the probability that the number of aftershocks is 
strictly positive, 
\begin{equation}
\langle\Delta^1\rangle = \langle\Delta\rangle \big/ \text{Pr}\{R>0\} ~,
\end{equation}
where $\text{Pr}\{R>0\}$ is the probability that the number $R$ of aftershocks is positive.
Accordingly, one introduce the mean rate $\nu^1$ of the non-empty clusters equal to
\begin{equation}
\nu^1 = \nu \big/ \text{Pr}\{R>0\} ~,
\end{equation}
where $\nu$ is the mean rate of noise earthquakes.

Obviously, $\text{Pr}\{R>0\}$ is given by
\begin{equation}
\text{Pr}\{R>0\} =  1 - p_1(0)~ ,
\end{equation}
where $p_1(0)$ is the probability that a noise earthquake does not
trigger any first generation aftershock at all. Using the parameterization
defined in the Appendix for the Hawkes model, 
$p_1(0)$ is given by expression \eqref{ponezone}, leading to
\begin{equation}
\nu^1 = \nu \cdot(n-\kappa)  \qquad \text{and} \qquad \langle\Delta^1\rangle =\frac{\overline{F}_\infty(n,\kappa,\alpha,1)}{n-\kappa}~ .
\end{equation}
In particular, in the critical case $n=1$, the mean duration of the 
non-empty clusters is given by
\begin{equation}\label{deltaonexpr}
\langle\Delta^1\rangle = \frac{1-\kappa (2-\alpha)}{(2-\alpha) \kappa (1-\kappa)}~ .
\end{equation}
As an example, taking $\kappa = 0.25$ and $\alpha = 1.5$ yields 
a mean duration of the non-empty clusters in the critical case equal to 
$\langle\Delta^1\rangle\simeq 9.33$. Recall that the unit time is the 
characteristic decay time of the Omori law $f_1(t)$ (\ref{expdisdef}). 

This allows us to define regimes of low seismicity as characterized by 
the following inequalities
\begin{equation}\label{lowsesmineq}
\nu \cdot  \langle\Delta\rangle \ll 1 \qquad \Leftrightarrow \qquad  \nu^1 \cdot \langle\Delta^1\rangle \ll 1~ ,
\end{equation}
which means that clusters are well individualized, being
separated by comparatively long quiet time intervals.

It is useful to generalize the mean duration $\langle\Delta^1\rangle$ of clusters
that contain at least one aftershock to the mean durations $\langle\Delta^m\rangle$ 
of the clusters that contain at least $m$ aftershocks. We now derive the general
equation allowing one to calculate these  $\langle\Delta^m\rangle$'s.
Using the total probability formula, one can represent 
the pdf $w(\varrho)$ of clusters durations in the form
\begin{equation}
w(\varrho) = p(0)\delta(\varrho)+\sum_{j=1}^\infty p(j) w(\varrho|j) ~,
\end{equation}
where $p(j)$ is the probability that a given noise earthquake 
triggers $j$ aftershocks of all generations, and $w(\varrho|j)$ is the conditional pdf of 
cluster durations under the condition that the number of aftershocks is equal to $j$. 
Accordingly, the pdf $w(\varrho|j\geqslant m)$ of the durations of the clusters that
have $m$ or more aftershocks is equal to
\begin{equation}
w(\varrho|j\geqslant m) = \frac{\displaystyle w(\varrho)- \sum_{j=1}^{m-1} p(j)w(\varrho|j)}{\displaystyle 1- \sum_{j=0}^{m-1} p(j)} ~.
\end{equation}
The corresponding conditional expectation $\langle \Delta^m\rangle$ is equal to
\begin{equation}
\langle \Delta^m\rangle = \int_0^\infty \varrho ~w(\varrho|j\geqslant m) d\varrho = \frac{\displaystyle \langle \Delta\rangle- \sum_{j=1}^{m-1} p(j)\langle \Delta|j\rangle}{\displaystyle 1- \sum_{j=0}^{m-1} p(j)}~ .
\end{equation}

\begin{figure}
\begin{center}
  \includegraphics[width=0.8\linewidth]{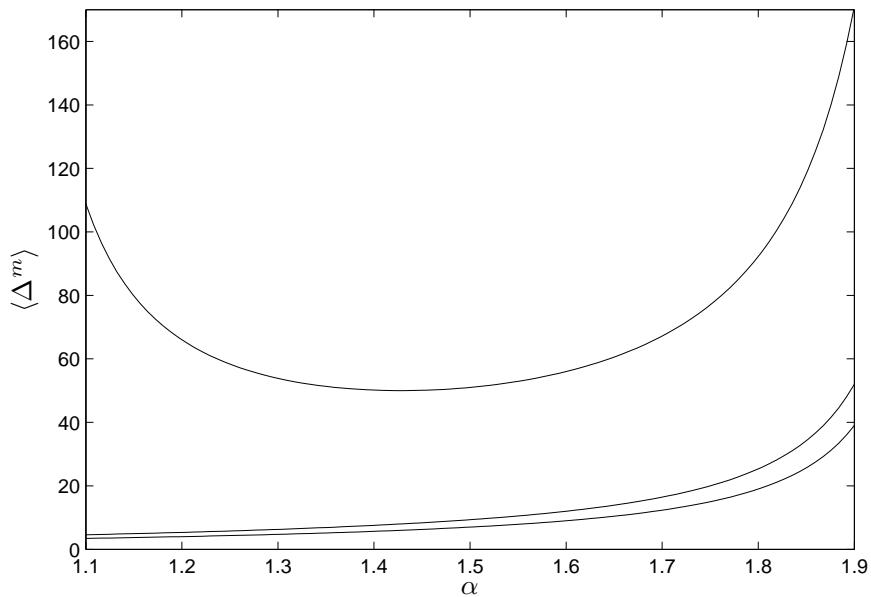}\\
\end{center}
  \caption{Dependences of the cluster durations $\langle \Delta\rangle$ \eqref{angledenone} (bottom curve), $\langle \Delta^1\rangle$ \eqref{deltaonexpr} (middle curve) and $\langle \Delta^2\rangle$ \eqref{delta2expr} 
(top curve) as functions of the exponent $\alpha$. The plot corresponds to the critical case $n=1$ with $\kappa=0.25$. Recall that $\langle \Delta^m \rangle$ is defined as the mean duration
of the clusters that contain at least $m$ aftershocks. In our notations, $\langle \Delta\rangle$
corresponds formally to $\langle \Delta^0 \rangle$, i.e., it also takes into account
the empty clusters for which the noise earthquake does not trigger any aftershock.
}\label{deltas}
\end{figure}

In particular, taking into account that
\begin{equation}\label{pwjexpr}
\begin{array}{c}
p(0) = p_1(0) = 1-n+\kappa , \qquad p(1) = p_1(1) = n - \alpha \kappa ,
\\[1mm] \displaystyle
w(\varrho|1) = f_1(\varrho) \qquad \Rightarrow \qquad \langle \Delta|j\rangle = 1~,
\end{array}
\end{equation}
we obtain
\begin{equation}\label{delta2exprbis}
\langle \Delta^2\rangle  = \frac{\langle \Delta\rangle -p(1)}{1-p(0)-p(1)} .
\end{equation}
Using \eqref{pwjexpr} for $n=1$ and the expression \eqref{angledenone}
for $\langle\Delta\rangle$, we obtain the mean duration of clusters containing more than one aftershock:
\begin{equation}\label{delta2expr}
\langle \Delta^2\rangle = \frac{1-\kappa(2-\alpha) (2-\alpha \kappa)}{\kappa^2 (\alpha-1) (2-\alpha)} ~.
\end{equation}
The dependences of the cluster durations $\langle \Delta\rangle$ \eqref{angledenone}, $\langle \Delta^1\rangle$ \eqref{deltaonexpr} and $\langle \Delta^2\rangle$ \eqref{delta2expr} as functions of 
the exponent $\alpha$ are shown in figure~\ref{deltas}.
Note the large jump in mean durations of clusters containing at least two aftershocks
compared with clusters containing at least one aftershock.

\section{Pdf of recurrence intervals}

\subsection{General relations}

The knowledge of the exact probability $\text{P}(\tau)$ \eqref{probsztauexpr}
allows one to calculate exactly the pdf $f(\tau)$ of the random 
waiting times $\{T_k\}$ between subsequent earthquakes. Indeed, a
general result of the theory of point processes states that
\begin{equation}\label{pdfpsidder}
f(\tau) = \langle\tau\rangle \frac{d^2 \text{P}(\tau)}{d\tau^2} ~,
\end{equation}
where $\langle\tau\rangle = \text{E}\left[T_k\right]$ denotes the mean waiting time between subsequent earthquakes. Therefore, the complementary cumulative distribution function (ccdf) of 
the random waiting times is equal to
\begin{equation}\label{lpsifirder}
\Psi(\tau) = \text{Pr}\left\{T>\tau\right\} = - \langle\tau\rangle  \frac{d P(\tau)}{d\tau} ~.
\end{equation}
By normalization, $\Psi(0)\equiv 1$, so that
\begin{equation}\label{meantrhudpdt}
 \frac{1}{\langle\tau\rangle} = - \frac{dP(\tau)}{d\tau}\bigg|_{\tau=0}~ .
\end{equation}

Using expressions  \eqref{probsztauexpr}, \eqref{prtauin} and \eqref{mathafoverline}, we have
\begin{equation}
\text{P}(\tau) = e^{-\nu \overline{F}(\tau) - \nu \tau}~.
\end{equation}
Making explicit $\overline{F}(\tau)$ with expression \eqref{overfexpr} yields
\begin{equation}
\frac{d P(\tau)}{d\tau} = \nu \frac{1-n \rho(\tau)+\kappa \rho^{\alpha}(\tau) }{1-n +\kappa \rho^{\alpha-1}(\tau)} e^{-\nu \overline{F}(\tau) - \nu \tau}~ .
\end{equation}
Using \eqref{meantrhudpdt}, we have
\begin{equation}\label{upstaumeanexpr}
\langle\tau\rangle = \frac{1-n}{\nu} \qquad \Rightarrow \qquad \Psi(\tau) = 
-\frac{1-n}{\nu} ~ \frac{d P(\tau)}{d\tau}~,
\end{equation}
and finally obtain
\begin{equation}\label{ccdfinterevent}
\Psi(\tau) = (1-n)  \frac{1-n \rho(\tau)+\kappa \rho^{\alpha}(\tau) }{1-n +\kappa \rho^{\alpha-1}(\tau)}  e^{-\nu \overline{F}(\tau) - \nu \tau} ~.
\end{equation}
Differentiating this last expression (\ref{ccdfinterevent}) with respect to $\tau$ 
yields the pdf $f(\tau)$ of waiting times between successive earthquakes:
\begin{equation}\label{pdftaugenexpr}
f(\tau) = \Phi(n,\kappa,\alpha,\rho(\tau),\nu)~,
\end{equation}
where
\begin{equation}\label{Phirenormexpr}
\begin{array}{c} \displaystyle
\Phi(n,\kappa,\alpha,\rho,\nu) =
\\[4mm] \displaystyle
\left[\mathcal{A}(n,\kappa,\alpha,\rho) + \nu \cdot \mathcal{B}(n,\kappa,\alpha,\rho) \right]  ~ e^{-\nu \left(\overline{F}(n,\kappa,\alpha,\rho)+\tau\right)} ,
\end{array}
\end{equation}
and
\begin{equation}\label{matkexpr}
\begin{array}{c} \displaystyle
\mathcal{A}(n,\kappa,\alpha,\rho) = (1-n) (1-\rho) \times
\\[5mm] \displaystyle
\frac{n (1-n)+ \kappa (\alpha-1 +(2 n-\alpha) \rho) \rho^{\alpha-2} -\kappa^2 \rho^{2(\alpha-1)}} {(1-n+\kappa \rho^{\alpha-1})^2} ,
\\[5mm] \displaystyle
\mathcal{B}(n,\kappa,\alpha,\rho) = (1-n) \left(\frac{1-n
\rho+\kappa \rho^\alpha}{1-n+ \kappa \rho^{\alpha-1}} \right)^2~ .
\end{array}
\end{equation}
Recall that $\rho$ represents $\rho(t)$, which is defined by expression (\ref{syjiukoiloyi}).

In the following subsections, we analyze in detail expressions \eqref{Phirenormexpr} and \eqref{matkexpr}
in order to derive the properties of the pdf $f(\tau)$ \eqref{pdftaugenexpr}.
 Expression \eqref{Phirenormexpr} suggests that it is natural to decompose the analysis
of $f(\tau)$ into two discussions, one centered on the term surviving in the limit $\nu \to 0$
and the other one.  The two next subsections analyze these two terms in turn.

\subsection{Case $\nu \to 0$}

Taking the limit $\nu \to 0$ amounts to neglecting the occurrence of any
noise earthquake within the window $(t,t+\tau)$ of analysis. 
As shown in the next subsection, this first case already reveals interesting properties
of the pdf $f(\tau)$, which remain valid in the general case $\nu> 0$.

Putting $\nu=0$ in expression \eqref{Phirenormexpr} and using \eqref{pdftaugenexpr}, we have
\begin{equation}\label{ftauthrua}
f(\tau) = \mathcal{A}\left[n,\kappa,\alpha,\rho(\tau)\right]  \qquad (\nu = 0)~ ,
\end{equation}
where the function $\mathcal{A}(n,\kappa,\alpha,\rho)$ is given by expression \eqref{matkexpr}.

The main asymptotics of the function $\mathcal{A}(n,\kappa,\alpha,\rho)$  are respectively
\begin{itemize}
\item at $\rho\ll 1$:
\begin{equation}\label{arholessone}
\mathcal{A}(n,\kappa,\alpha,\rho) \simeq \kappa (1-n) (\alpha-1) ~ \frac{\rho^{\alpha-2}}{(1-n+\kappa \rho^{\alpha-1})^2} ~, \qquad  \rho\ll 1~.
\end{equation}
This regime $\rho\ll 1$ corresponds to $\tau \ll 1$ and thus 
$\rho \simeq \tau$. Relation \eqref{arholessone} thus leads to
\begin{equation}\label{ftaulessonetwo}
f(\tau) \simeq  \kappa (1-n) (\alpha-1) ~ \frac{\tau^{\alpha-2}}{(1-n+\kappa \tau^{\alpha-1})^2} , \qquad \tau \ll 1~ .
\end{equation}

\item At $\rho\to 1$:
\begin{equation}\label{asymprtone}
\mathcal{A}(n,\kappa,\alpha,\rho) \simeq \mathcal{C} (1-\rho) ~, \qquad 1-\rho \ll 1 , \qquad \mathcal{C} = \frac{(1-n) (n-\kappa)}{1-n+\kappa} ~.
\end{equation}
This second asymptotic $\rho\to 1$ corresponds to  $1-\rho\ll 1$,
which is equivalent to the condition $\tau\gg 1$.
Using expression \eqref{atauexp},  relation \eqref{asymprtone} leads to
\begin{equation}\label{ftauexpasymp}
f(\tau) \simeq \mathcal{C} ~ e^{-\tau} , \qquad \tau \gg 1 .
\end{equation}
\end{itemize}

The denominator of expression (\ref{ftaulessonetwo}) determines a new characteristic 
time scale
\begin{equation}\label{taustardef}
\tau_c = \left(\frac{1-n}{\kappa}\right)^{\frac{1}{\alpha-1}} ~,
\end{equation}
from which one can define a critical value for the branching ratio (\ref{wryjujiuk})
\begin{equation}
n_c = 1-\kappa~,
\label{ehyhyt}
\end{equation}
such that $\tau_c > 1$ for $n < n_c$ and $\tau_c < 1$ for $n > n_c$.
We shall refer to the case $n<n_c$ as the \emph{subcritical} regime, while
$n_c<n<1$ is called the \emph{near-critical} regime.

For $n < n_c$ (subcritical regime),  $\tau_c\gtrsim 1$, 
and one may replace the asymptotics \eqref{ftaulessonetwo} by the pure power law:
\begin{equation}\label{onepowerasym}
f(\tau) \simeq \kappa ~ \frac{\alpha-1}{1-n} ~ \tau^{-(2-\alpha)}~, \qquad  n \lesssim n_c~.
\end{equation}

In contrast, in the near-critical case $n_c<n<1$,
expression \eqref{ftaulessonetwo} leads to two power asymptotics:
\begin{equation}\label{twopowerasym}
f(\tau) \simeq
\begin{cases} \displaystyle
\kappa ~ \frac{\alpha-1}{1-n} ~ \tau^{-(2-\alpha)}~ , & \tau \ll \tau_c ~,
\\[4mm] \displaystyle
(1-n) \frac{\alpha-1}{\kappa}~ \tau^{-\alpha}~ , & \tau_c \ll \tau \ll 1~ ,
\end{cases}
\qquad n_c < n <1~ .
\end{equation}

\begin{figure}
\begin{center}
  \includegraphics[width=0.8\linewidth]{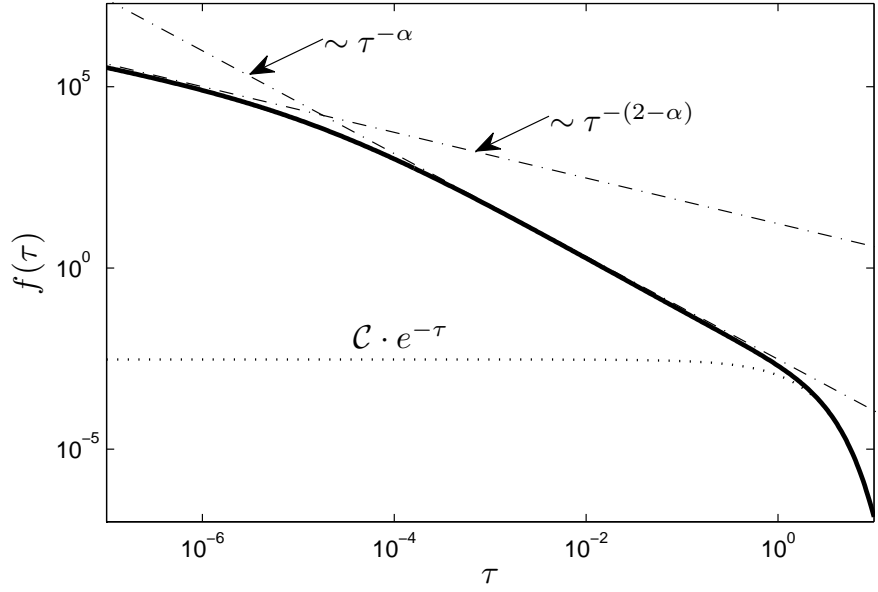}\\
\end{center}
  \caption{Solid line: Plot of the pdf $f(\tau)$ \eqref{ftauthrua} 
of waiting times between successive earthquakes
for $\nu=0$, $\kappa=0.25$, $\alpha=1.5$ and $n=0.999$. 
The dotted lines show the power law asymptotics \eqref{twopowerasym} and the exponential 
asymptotic behavior \eqref{ftauexpasymp}.}\label{nuzero999}
\end{figure}

\begin{figure}
\begin{center}
  \includegraphics[width=0.8\linewidth]{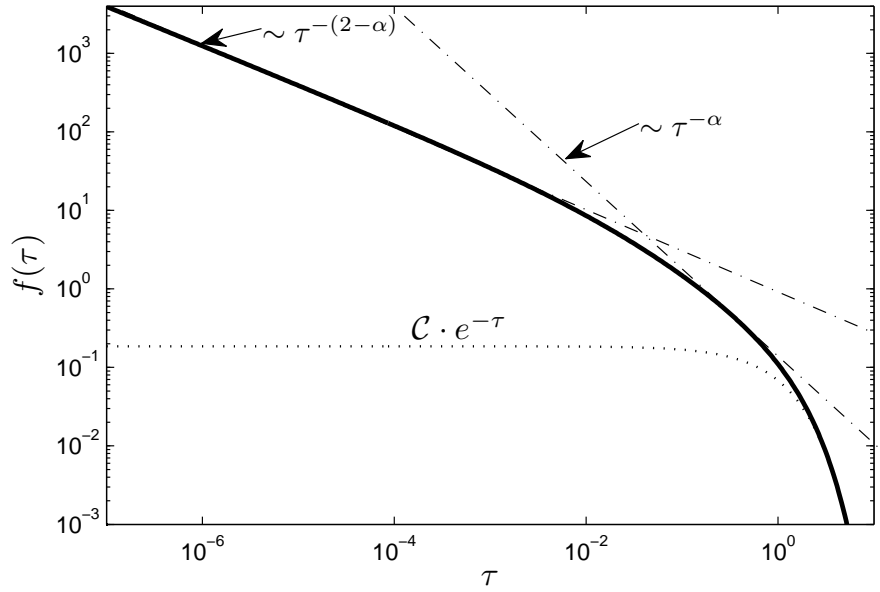}\\
\end{center}
  \caption{Same as figure \ref{nuzero999} except for the value of the branching ratio $n=0.9$.}\label{nuzero9}
\end{figure}

Figure~\ref{nuzero999} shows the pdf $f(\tau)$ \eqref{ftauthrua}
of the waiting times between successive earthquakes for
$\nu=0$, $\alpha=1.5$, $\kappa=0.25$ ($n_c=0.75$), and in the near-critical case $n=0.999$. 
The two power law asymptotics \eqref{twopowerasym} and the exponential asymptotics 
\eqref{ftauexpasymp} for  large $\tau$'s are clearly visible. 
Figure~\ref{nuzero9} is the same as figure ~\ref{nuzero999}, except 
for the value $n=0.9$. Although, formally, this value belongs also to the near-critical case
($n=0.9 > n_c =0.75$),  the intermediate power law asymptotics $\tau^{-\alpha}$ 
is barely visible and, for any $\tau\ll 1$, the subcritical power asymptotics $\tau^{-(2-\alpha)}$ dominates
at short times.

\begin{figure}
\begin{center}
  \includegraphics[width=0.8\linewidth]{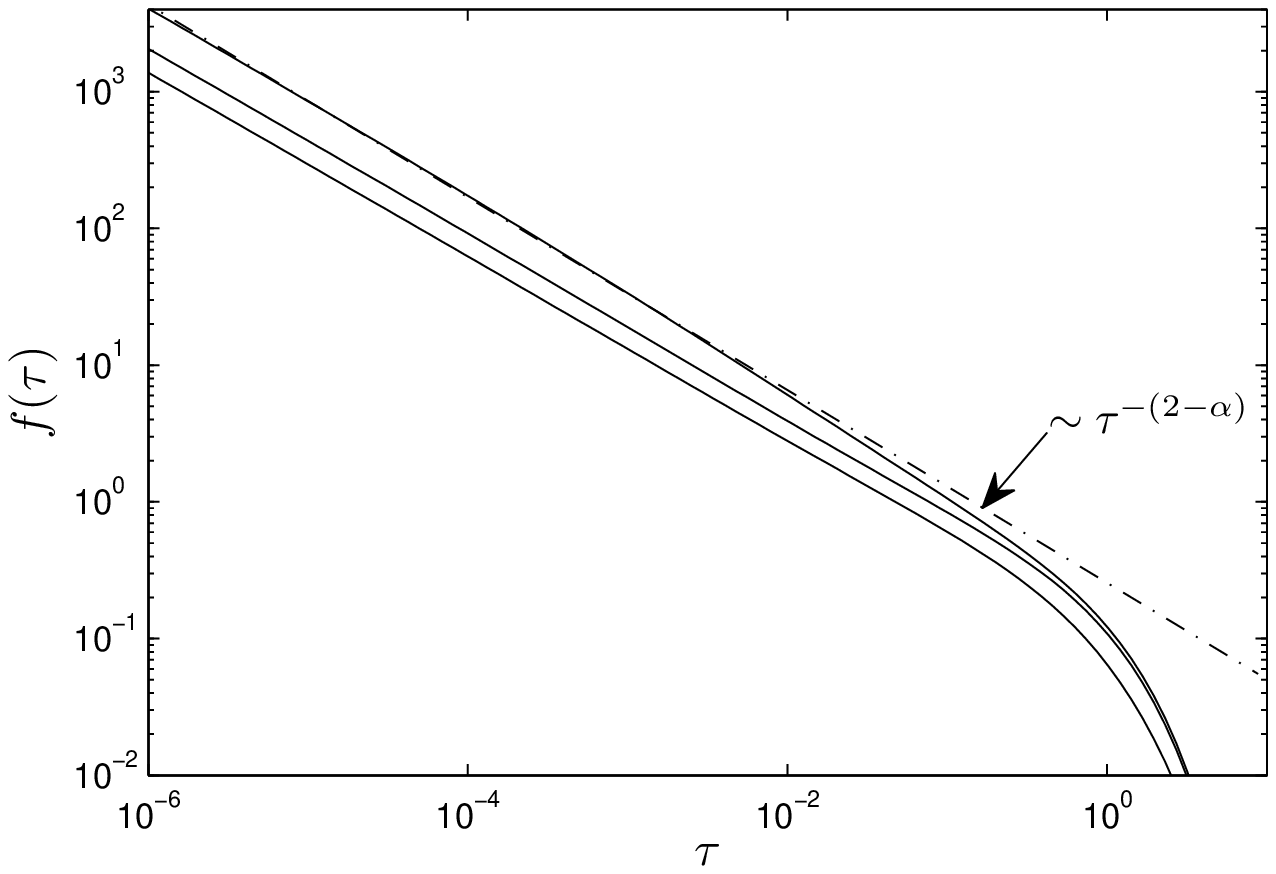}\\
\end{center}
  \caption{Solid line: Log-log plot of the pdf $f(\tau)$ \eqref{ftauthrua} for $\nu=0$, $\kappa=0.25$, $\alpha=1.5$ and for three $n$ values in the subcritical regime. Top to bottom: $n=0.8; 0.6; 0.4$. The dotted straight line is the subcritical power law asymptotics \eqref{onepowerasym}.}\label{nuzeroens}
\end{figure}

Figure~\ref{nuzeroens} shows log-log plots of the pdf $f(\tau)$ for $\nu=0$, $\kappa=0.25$, $\alpha=1.5$ and different values of the branching ratio $n$ belonging to the subcritical regime.
The subcritical power law asymptotics $\tau^{-(2-\alpha)}$ is 
dominant and the different pdf's are similar.

\subsection{General case $\nu \neq 0$}

We now take into account the occurrence of
noise earthquakes within the window $(t,t+\tau)$ of analysis. 
The main qualitative difference between this case and the previous one $\nu =0$ is
the replacement of the large $\tau$ asymptotics \eqref{ftauexpasymp} by
\begin{equation}
f(\tau) \simeq (1-n) \left( \frac{n-\kappa}{1-n+\kappa}  e^{-\tau} + 
\nu\right) e^{-\nu (\langle \Delta \rangle +\tau)} , \qquad \tau \gg 1 ~.
\label{ryjrujkiku}
\end{equation}

\begin{figure}
\begin{center}
  \includegraphics[width=0.8\linewidth]{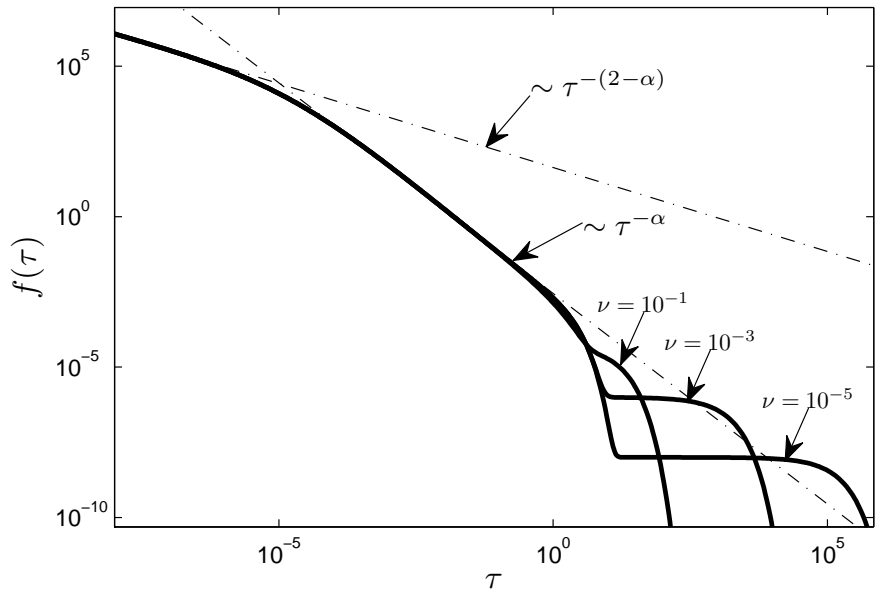}\\
\end{center}
  \caption{Solid line: Loglog plots of the recurrence times pdf $f(\tau)$ \eqref{pdftaugenexpr} in the case $\kappa=0.25$, $\alpha=1.5$  $n=0.999$ and three subcritical $\nu$ values: $\nu=0.1; 10^{-3}$ and $\nu=10^{-5}$. Dotted straight lines are the subcritical and near-critical power asymptotics \eqref{ftauexpasymp}}\label{nu999}
\end{figure}

The interesting regime of well-defined clusters occurs for $\nu\ll 1$,
for which the typical waiting time between noise earthquakes is much larger
than the characteristic decay time of the Omori law, which is the typical waiting time
between a noise earthquake and its aftershocks.
In the interval $1\ll \tau \ll 1\big/\nu$, expression (\ref{ryjrujkiku}) reduces to (\ref{ftauexpasymp})
and this exponential decay can be replaced by the plateau
\begin{equation}\label{plateau}
f(\tau) \simeq (1-n) \nu~ , \qquad 1\ll \tau \ll 1\big/\nu~ .
\end{equation}
For $\tau\gtrsim 1/\nu$, expression (\ref{ryjrujkiku}) simplifies into
\begin{equation}\label{nuexpasymp}
f(\tau) \simeq (1-n) \nu  e^{-\nu \langle \Delta \rangle} ~e^{-\nu \tau} , \qquad \tau \gtrsim 1/\nu ~.
\end{equation}
These different regimes are illustrated in
figure~\ref{nu999}, which depicts the pdf of waiting times between successive earthquakes
for $\kappa=0.25$, $\alpha=1.5$, $\nu=0.999$ and three values of 
the positive parameter $\nu=10^{-1}; 10^{-3}; 10^{-5}$.
One can clearly observe the subcritical and near-critical power law asymptotics \eqref{twopowerasym},
as well as the plateau \eqref{plateau} joining the exponential asymptotics \eqref{ftauexpasymp} and \eqref{nuexpasymp}. The plateau is especially visible for the smallest values 
$\nu=10^{-3}$ and $\nu=10^{-5}$.

\section{Conclusion}

In this paper, we have studied the impact of the power law 
form \eqref{poneralasymp} of the distribution of fertilities on the distribution of the recurrence times.
In contrast with previous studies, we have considered the simple case
where the Omori law is not heavy-tailed but is described by an exponential function.
Motivated by real applications, this choice allows us to develop an exact analytical treatment
using the formalism of generating probability functions.

Our analysis emphasizes the importance of three time scales controlling the different
regimes of the probability density function (pdf) $f(\tau)$ of waiting times between 
successive earthquakes:
\begin{itemize}
\item the characteristic decay time of the exponential Omori law $f_1(t)$ (\ref{expdisdef})
describing the occurrence of first generation aftershocks,
taken as our time unit, 
\item the average waiting time $1/\nu$ between two successive ``noise earthquakes'', 
which constitute the exogenous sources of the self-excited processes followed by their aftershocks,
\item the characteristic time $\tau_c$ defined by expression (\ref{taustardef}) 
associated with the self-excited cascades of aftershocks.
\end{itemize}

In the interesting and relevant regime where earthquake clusters
are well defined, namely when the typical waiting time till the first aftershocks (unit time) is smaller than
the waiting times $1/\nu$ between noise earthquakes, we have found that the pdf of recurrence times
exhibits several intermediate power law asymptotics:
\begin{enumerate}
\item For $\tau \ll \tau_c$, $f(\tau) \sim \tau^{-(2-\alpha)}$.
\item For $\tau_c \ll \tau \ll 1$, $f(\tau) \sim \tau^{-\alpha}$.
\item For $1 \ll \tau \ll 1/\nu$, $f(\tau) \simeq (1-n) \nu = const$.
\item For $1/\nu \lesssim \tau$, $f(\tau) \simeq e^{-\nu \langle \Delta \rangle} ~e^{-\nu \tau} $.
\end{enumerate}
In these formulas, $\alpha$ is the exponent of the power law distribution
of fertilities  $p_1(r) \sim r^{-\alpha-1}$ (\ref{poneralasymp}), which is 
the pdf of the number of first generation aftershocks triggered by a given event  of any type.
In turn, $n$ stands for the branching ratio defined by equation (\ref{wryjujiuk}),
i.e. the average number of daughters of first generation per mother event, and
$\langle \Delta \rangle$ is the mean duration of a cluster that starts with a noise earthquake
and ends with its last aftershock over all generation. It is given by expression 
(\ref{nglDelgen}).

Only the first two intermediate asymptotics  $f(\tau) \sim \tau^{-(2-\alpha)}$ and 
$f(\tau) \sim \tau^{-\alpha}$ at short time scales reflect
the influence of the power law distribution
of fertilities (\ref{poneralasymp}), which is revealed by the remarkable
effect of the cascade of triggering over the population of aftershocks of 
many different generations.

Finally, let us stress the differences between the present investigation
and our previous work \cite{SaiSor2006,SaiSor2007}
on the same problem.
In Refs.~\cite{SaiSor2006,SaiSor2007}, we determined 
the asymptotic behavior at long times of the distribution of recurrence times,
under the approximation that it was sufficient to consider only one
aftershock at most per mother event (`noise earthquake' in the present
terminology).  In addition, we considered the standard power law Omori law 
(\ref{omorilawexpr}) and not the exponential law (\ref{expdisdef}).
By performing a detailed analysis made possible by the 
use of the exactly tractable exponential Omori law (\ref{expdisdef}),
the present paper has thus demonstrated the existence of additional
short-time intermediate asymptotics that reveal the distribution of
fertilities. This opens the possibility to estimate the exponent $\alpha$
of the distribution of cluster sizes from purely dynamic measures
of activity.

\clearpage

\section*{Appendix: Statistics of first generation aftershocks}

Given the power law \eqref{poneralasymp} for the right tail of the pdf $\{p_1(r)\}$
of the number $R_1$ of the first generation aftershocks  probabilities, 
we show that the leading relevant terms of the expansion of the GPF $G_1(z)$ 
in powers of $(1-z)$ take the form  
\begin{equation}\label{gonepowgamdef2}
G_1(z) = 1 - n (1-z) + \kappa (1-z)^\alpha , \qquad \alpha\in(1,2] ,
\end{equation}
so that the corresponding auxiliary function $Q(y)$ \eqref{qytrugoneomz} takes the form
\begin{equation}\label{qyalphadef2}
Q(y) = 1 - n y + \kappa y^\alpha .
\end{equation}
They depend on the branching ratio $n$ defined in (\ref{wryjujiuk}), the 
exponent $\alpha$ of the power law distribution \eqref{poneralasymp}
of the number of first-generation aftershocks. The additional scale parameter
$\kappa$ satisfies the following inequalities
\begin{equation}\label{gamkapenineq}
\begin{cases}
0 < \alpha \kappa < n , & n\leqslant 1~ ,
\\
n-1 < \alpha \kappa < n , & n > 1~ ,
\end{cases}
\end{equation}
which ensure the necessary constraints
\[
0\leqslant p_1(0) \leqslant 1 , \qquad \text{and} \qquad 0\leqslant p_1(1) \leqslant 1 .
\]

Rather than deriving the form (\ref{gonepowgamdef2}) from \eqref{poneralasymp},
it is more convenient to show that the tail of the pdf $\{p_1(r)\}$ whose GPF is 
given by (\ref{gonepowgamdef2}) is the power law \eqref{poneralasymp}.
Given (\ref{gonepowgamdef2}) and the definition linking $G_1(z)$ to $p_1(r)$,
namely $G_1(z)= \sum_{j=0}^{+\infty} p_1(r) z^r$, we obtain 
\begin{equation}\label{ponezone}
p_1(0) = 1- n + \kappa , \qquad p_1(1) = n -\alpha \kappa ~,
\end{equation}
and
\begin{equation}\label{pkmoretwo}
\begin{array}{c} \displaystyle
p_1(r) = \kappa (-1)^k \binom{\alpha}{r} = \frac{\kappa (-1)^r \Gamma(\alpha+1)}{\Gamma(r+1) \Gamma(\alpha-r+1)} ,
\\[4mm]\displaystyle
r \geqslant 2 , \qquad \alpha \in (1,2) .
\end{array}
\end{equation}

Using the properties of gamma functions and in particular the well-known equality
\begin{equation}
\Gamma(z) \Gamma(1-z) = \frac{\pi}{\sin\pi z}~ ,
\end{equation}
we obtain that
\begin{equation}
(-1)^r \Gamma(\alpha-r+1) = \frac{\pi}{\Gamma(r-\alpha) \sin[\pi(\alpha-1)]} , \qquad \alpha\in(1,2) .
\end{equation}
Accordingly, expression \eqref{pkmoretwo} takes the form
\begin{equation}\label{pkgamevent}
p_1(r) = c \cdot\frac{\Gamma(r-\alpha)}{\Gamma(r+1)} , \qquad c := \frac{\kappa}{\Gamma(-\alpha)}  .
\end{equation}
using the asymptotic relation
\[
\frac{\Gamma(r+a)}{\Gamma(r+b)} \simeq r^{a-b} , \qquad r\to\infty ,
\]
we finally recover the power law \eqref{poneralasymp}.
 
The case where the exponent $\alpha=2$ in expression (\ref{gonepowgamdef2}) of 
$G_1(z)$ requires a special mention. Indeed, this form describes the 
special situation in which each noise earthquake (and any aftershock as well)
can trigger not more than two first generation aftershocks. 
Accordingly, there are, in general, only three nonzero probabilities
\begin{equation}
p_1(0) = 1-n+\kappa , \qquad p_1(1) = n- 2 \kappa , \qquad p_1(2) = \kappa  \qquad (\alpha=2)~ .
\end{equation}
This special situation arises due to the fact that the GPF $G_1(z)$ has been
truncated beyond the quadratic order $(1-z)^2$. It will not be considered further in this paper.

\end{document}